\documentclass[proof]{WileyASNA-v1}

\articletype{Article Type}%
\usepackage{pgfplots}
\usetikzlibrary{calc}
\pgfplotsset{compat=newest}

\colorlet{myblue}{blue!30}

\usepackage{tikz}
\usepackage{graphicx}
\usepackage{bigints}
\usepackage{pgfplots}
\usetikzlibrary{calc}
\pgfplotsset{compat=newest}

\colorlet{myblue}{blue!30}

\articletype{Article Type}%

\received{<day> <Month>, <year>}
\revised{<day> <Month>, <year>}
\accepted{<day> <Month>, <year>}

\usepackage{mathtools}
\usepackage{amsmath}
\usepackage{palatino}
\newdimen\temp
\temp=\abovedisplayskip
\usepackage{blindtext}
\begin{document}

\title{Causality and the Arrow of Time in the Branch-Cut Cosmology}%\protect\thanks{This is an example for title footnote.}}

\author[1]{Benno Bodmann}

\author[2,3]{C\'esar A. Zen Vasconcellos*}

\author[4]{Jos\'e de Freitas Pacheco}

\author[5,6]{Peter O. Hess}

\author[2]{Dimiter Hadjimichef}

\authormark{Benno Bodmann \textsc{et al}}

\address[1]{Universidade Federal de Santa Maria (UFSM), Santa Maria, Brazil}

\address[2]{Instituto de F\'isica, Universidade Federal do Rio Grande do Sul (UFRGS), Porto Alegre, Brazil}

\address[3]{International Center for Relativistic Astrophysics Network (ICRANet), Pescara, Italy}

\address[4]{Observatoire de la C\^ote d'Azur, Nice, France}

\address[5]{Universidad Nacional Aut\'onoma de Mexico (UNAM), M\'exico City, M\'exico}

\address[6]{Frankfurt Institute for Advanced Studies (FIAS), Hessen, Germany}

\corres{*C\'esar A. Zen Vasconcellos. \email{cesarzen@cesarzen.com}}

%\presentaddress{Present address}

\abstract[Abstract]{We basis our initial analysis of the arrow of time on a relationship between the time evolution operator of quantum system and the time-independent density operator which describes the equilibrium state of a many-particle system at temperature $T$. We highlight through this analysis the identification of the imaginary temporal component of the 
branch-cut complex cosmic form factor with the direction in which the time-parameter flows globally, or the arrow of time. 
As a novelty, in this work we calculate the number of branches in the branch-cut universe to achieve causality involving the global time of evolution of the universe and the local time of travel of the light around each Hubble horizon. The preliminary result obtained is comparable to 60 e-folds of contraction in the FLRW cosmic scale factor $a(t) $ to overcome causality achieved in the bouncing model.
}


\maketitle

%%%%%%%%%%%%%%%%%%%%%%%%%%%%%%%%%%%
\section{introduction}
%%%%%%%%%%%%%%%%%%%%%%%%%%%%%%%%%%%%

The arrow of time problem, the unidirectional flow of time embodied in the second law of thermodynamics, represents one of the main fundamental problems of physics, since the macroscopic time-asymmetry emerges from a symmetrical-time micro-physics~\citep{Eddington, Ellis}.

The meaning of time and the direction and passage of time, is a mystery that has challenged our understanding and imagination since the dawn of our civilization, which become strikingly illustrated in a thought of St. Augustine of Hippo, a known philosopher and theologian: 
“If nobody asks me, I know what time is, but if I am asked then I am at a loss what to say”.  In a brief synthesis on the philosophical journey of humanity regarding the meaning of time, Parmenides denied time’s existence, Plato imagined a world of ideas outside it and Hegel spoke of the moment in the spirit transcends temporality~\citep{Reichenbach}.

Albert Einstein, presumably inspired by Parmenides\footnote{According to Karl~\citet{Popper}.}$^, $\footnote{Parmenides’ deduction of the nature of reality led him to conclude “that reality [is], and must be, a unity in the strictest sense and that any change in it [is] impossible” and therefore  “the world as perceived by the senses is unreal”~\citep{SEP}.}, by conceiving the `Einstein Universe'~\citep{Einstein1917},  paved his view about the universe, a four-dimensional space-time ball with fixed radius, in which time has no beginning, being infinite in both directions, an universe that has always existed and will always exist, known nowadays as the `Block Universe' (BU). 

In Einstein's conception, the universe was assumed to be a continuous, immutable, four-dimensional space-time block that contains all moments of time, without attributing to them any special particularity, in which the past, the present and the future  are equivalent and indistinguishable attributes. In this sense, there is no `now' that can be uniquely called the present, no flow of time and not even evolution of space-time. This representation implicitly embodies the idea that time is just an illusion~\citep{Barbour2000}, emphasized once by Albert Einstein that the true reality is timeless\footnote{Albert Einstein in a letter to the family of Michele Besso his collaborator and closest friend, once wrote:
``Now he has departed from this strange world a little ahead of me. That means nothing. People like us, who believe in physics, know that the distinction between past, present, and future is only a stubbornly persistent illusion''~\citep{CHRISTIES2020}.}

In contrast to the BU view, George~\citet{Ellis,Ellis2014} advocates that the true nature of spacetime is best represented as an Emergent Block Universe (EBU). The  main feature of the EBU is an indefinite spacetime, ``which grows and incorporates ever more events, `concreting' as time evolves along each world line, with quantum uncertainty continually changing to classical definiteness''. In this view, the present represents a boundary that separates the past, which once existed, from the future, which does not yet exist and is indeterminate because of quantum uncertainties and the arrow of time is distinguished from the direction of time, which is non-locally defined in the evolving block universe context\footnote{\citet{Ellis} distinguishes the concepts of "direction of time" and the "arrow of time" as follows: the direction of time corresponding to the cosmological determined direction in which time flows globally and the arrow of time is the locally determined direction in which time flows at any time in the evolution of the universe. The first concept represents the way spacetime is continuously increasing as an Evolving Block Universe and the second represents the way physics and biology manifest the flow of time locally.}.

In the following we discuss an aspect that may shed some light for a better understanding of the connection between macro- and micro-physics and the origin of the arrow of time. Based on a relationship, not yet sufficiently understood, between the time evolution operator of quantum system and the time-independent density operator which describes the equilibrium state of a many-particle system at temperature T. As addressed in the abstract, we highlight in this contribution the identification of the arrow of time with the imaginary temporal component of the 
branch-cut complex cosmic form factor
through a Wick rotation which describes, in contrast to Ellis' view~\citep{Ellis}, the direction in which the parameter time flows globally. 
As a novelty, in this work we calculate the number of branches in the branch-cut universe to achieve causality involving the global time of evolution of the universe and the local time of travel of the light around each Hubble horizon. 

In the following, we discuss the conceptions about the arrow of time addressed by branched cosmology.

%%%%%%%%%%%%%%%%%%%%%%%%%%%
\section{The arrow of time in Branch-cut cosmology}
%%%%%%%%%%%%%%%%%%%%%%%%%%%

The arrow of time, a term coined by Arthur Stanley~\citet{Eddington} and identified as the macroscopic unidirectional flow of time, is one of the most enigmatic features of the evolutionary universe, challenging our understanding about. 
Encapsulated in the second law of thermodynamics\footnote{The second law of thermodynamics establishes that in an isolated system, entropy tends to increase with time 
and that entropy changes in the universe can never be negative.}, the asymmetrical thermodynamic macroscopic arrow of time emerges from a micro-physics symmetrical in time. This is the crucial aspect where the difficulty of understanding its origin resides.

The superposition principle of quantum mechanics establishes 
that a physical microscopical system, described by a general state vector, $|\Psi \rangle$, can be expanded as a linear superposition of $n$ unit orthogonal basis vectors, $|\phi_{n \hat{\cal O}}\rangle$, that describes the eigenstates of any quantum observable represented by the operator $\hat{\cal O}$
\begin{equation}
|\Psi \rangle = \sum_i^{\infty} a_n  |\phi_{n \hat{\cal O}}\rangle \, \quad \mbox{with} \quad a_n =    \langle \phi_{n \hat{\cal O}}|\Psi \rangle \, ,
\end{equation}
where the coefficients $a_n$ are, in general, functions of time.
Limiting ourselves to the non-relativistic realm, the time evolution of $|\Psi \rangle $ is described by the Schr\"odinger equation
\begin{equation}
i \hbar \frac{\partial}{\partial t} |\Psi\rangle  = \hat{H} |\Psi \rangle  \, ,
\end{equation}
where $\hat{H}$ represents the time-independent Hamiltonian. Quantum mechanics is therefore characterized, before a measurement, by  time-reversible (time-symmetrical) processes, not contemplating a defined temporal orientation, or time arrow.
However, in the measurement process, there is a probabilistic and time-irreversible (time-asymmetrical) reduction of the original state vector, in which the original projection of $|\Psi \rangle$ into a subspace of eigenvectors 
`collapses' to one of the eigenstates, causing the arrow of time to manifest at the quantum level\footnote{After the time-irreversible process of
state-vector reduction has taken place, the past emerges, with the previous quantum uncertainty replaced by the classical certainty of definite particle identities and states'~\citep{Ellis}.}. 

One aspect that may shed some light for a better understanding of the connection between macro- and micro-physics involves a relationship, not yet sufficiently understood, 
between the time evolution operator $\hat{U}(t)$ of a quantum system
and the statistical time-independent density operator, $\hat{U}(T)$: 
\begin{equation}
\hat{U}(t)
= e^{-i\hat{H}t/\hbar} \xRightarrow[]{t \to -i \tau} e^{-\hat{H} \tau /\hbar}  \xRightarrow[]{\tau/\hbar \to 1/T \equiv \beta} \hat{U}(T) = e^{-\beta \hat{H}} \, , 
\end{equation}
where the Euclidean (imaginary) time $\tau$ is cyclic with period $\beta = 1/T$.  $\hat{U}(t)$ governs how a quantum system evolves in time  and
$\hat{U}(T)$ describes the equilibrium statistical state of a many-particle system at a temperature T. Thus,
the understanding of this connection lies in the interpretation of the role of the Hamiltonian $\hat{H}$ in quantum mechanics and statistical mechanics and the meaning of the transformation $t \it \to -i\tau$. In quantum mechanics, as well as in quantum field theory,
the Hamiltonian $\hat{H}$ acts as the generator of the Lie group of time translations, while in statistical mechanics it operates as the Boltzmann weight in an ensemble. The Wick rotation transformation\footnote{Wick's rotation, in relating classical statistical mechanics and quantum field theory,  makes it possible to give a more comprehensible physical interpretation to the path integrals formalism when expressed in terms of a Euclidean space-time as opposed to Minkowski's space-time, such as overcoming ill-definitions and the presence of singularities.} $t \to -i\tau$, with $\tau \in \Re_+$, replaces the Minkowski real time $t$ with the Euclidean imaginary time $i \tau$ and makes the density operator of statistical mechanics to act as an imaginary time evolution operator, as long as the substitution 
$t \to \tau \equiv -i\hbar \beta = -i \hbar/T$ 
is made, giving rise to a correspondence between cyclic time and the temperature. 

In the context of the second law of thermodynamics, the expansion of the universe is associated with the thermodynamic arrow of time pointing from the past to the future. However, the thermodynamic arrow would reverse during a contraction phase of the universe or inside black holes~\citep{Hawking1985}.

Two scenarios have been outlined in the branch-cut cosmology.
The first scenario is characterized by a branch-point and a branch-cut, with the suppression of a primordial singularity, in which the universe continuously evolves from the negative imaginary cosmological-time sector, 
where the contraction of the universe takes place,
to the positive one, where the expansion of the universe occurs~\cite{Zen2020,Zen2021a,Zen2021b,Zen2022}.
%\cite{Zen2020}, \cite{Zen2021a}, \cite{Zen2021b}, \cite{Zen2022}.
  This scenario, due to the absence of a singularity, may be characterized by a transition region with dimensions that surpass Planck's dimensions according to Bekenstein's criterion~\citep{Bekenstein1981,Bekenstein2003}, separating the contraction and expansion phases~\citep{Zen2023a}. 

In the second scenario, the branch-cut and the branch-point disappear after realization of imaginary time through a Wick rotation, which is replaced here by the real and continuous thermal time (temperature)~\citep{Zen2020,Zen2021a,Zen2021b,Zen2022}. In this scenario, 
 the connection between the negative and positive imaginary cosmological-time sectors is broken as a result of the Wick rotation, and a mirrored parallel evolutionary universe adjacent to our own is nested in the fabric of space and time, with its evolutionary process oriented in the opposite direction to that corresponding to the positive imaginary-time sector~\citep{Zen2020,Zen2021a,Zen2021b,Zen2022}. 
 
 Singularity means there is no way for spacetime to start smoothly. Branch-cut cosmology, alternatively, proposes an absolutely non-temporal beginning in the imaginary sector, a configuration of pure space, through a Wick rotation that replaces the imaginary component of time with the temperature, the thermal time, that flows in the opposite direction of the arrow of time in the expansion phases of the first and second scenarios.
In the first scenario, in the contraction phase, before entering into the expansion phase, the temperature and entropy of the branch-cut universe must reach values consistent 
with the corresponding ones in the expansion phase. For this to happen, the temperature of the universe in the contraction phase must increase, but the entropy must decrease, as determined by thermodynamics, reversing this way the arrow of time.  
 In the contraction sector of the first scenario, as the transition region approaches, there occurs a progressive decrease in the entropy and an increase in the temperature of the universe, so there is a critical region, whose dimensions are determined by the Bekenstein Criterion~\citep{Zen2023a}, where the entropy reaches its minimum value and the temperature in contra-position its maximum value.

In order to maintain the ``past-to-future''  global orientation, we propose as a novelty a time arrow oriented towards the decreasing of entropy in the contraction sector of the universe, and the conventional conception in the expansion phase. This proposition would be valid for both scenarios of the branch-cut cosmology, as in the first scenario there is a contraction sector followed by an expansion phase of the universe, while in the second scenario all sectors are associated with the expansion of both, ours and the mirror-universe. In this conception, the cosmological arrow of time is determined as the direction in which ``time'', from the macroscopic point of view, flows globally. This conception differs from that of Ellis, who identifies the arrow of time with the locally determined direction in which time flows at any time in the evolution of the universe. In the branch-cut cosmology, in turn, this local component is identified with the corresponding real time component that results from the complexification of the FLRW metric ~\citep{Zen2021a}.

\section{Number of branches to achieve causality in the branch-cut cosmology}

The big bang model difficulty in explaining the observed homogeneity of causally disconnected regions of space, in the absence of a mechanism that establishes the same initial conditions everywhere, gave rise to the cosmological fine-tuning problem, or horizon problem, firstly pointed out by Wolfgang~\citet{Rindler}. 

The horizon problem 
arises because the patch corresponding to the observable universe was never causally connected in the past~\cite{Ijjas2014,Ijjas2018,Ijjas2019}.  In the present time ($t = t_0$), the patch size, $R(t_0)$ and the horizon size, $H^{-1}(t_0)$, are equal, ie, $R( t_0) = H^{-1}(t_0)$. In earlier times, the ratio between the horizon size to the patch size decreases monotonically extrapolating back in time as $a(t) \rightarrow 0$ in the form 
$a^{\epsilon}(t)/a(t)$:  
\begin{equation}
\frac{H^{-1}(t)}{a(t)} \sim 
\frac{a^{\epsilon}(t)}{a(t)} = a^{\epsilon - 1}(t) \quad
\mbox{and} \quad  \lim_{a(t) \rightarrow 0}{a(t)^{\epsilon - 1}} \rightarrow 0, \label{aepsilona}
\end{equation}
with the horizon size approaching zero faster than the patch size.
According to the CMB measurements, the density and temperature were almost uniform throughout the primordial patch (last CMB surface scattering). Explaining the uniformity of the CMB at length scales greater than the size of the horizon at the last scattering surface grounded on the CMB measurements
and at all previous times fundamentally constitutes the horizon problem. Moreover, the CMB measurements also reveal a spectrum of small amplitude density fluctuations, nearly scale-invariant whose explanation constitutes the inhomogeneity problem~\cite{Ijjas2014,Ijjas2018,Ijjas2019}.

This non-causal behavior of the patch size and the horizon size, --- since their ratio continuously decreases when extrapolating $a(t)$ backwards in time ---, represents one of the fundamental limitations of standard cosmology. Combined with the primordial singularity, where any trace of causality completely disappears, these two factors represent the main roots for solving the problems of the cosmic singularity, horizon, inhomogeneity and flatness of the universe. The 
smoothness problem of the universe is related in turn to the
{\it cosmic anisotropy factor}, $\Omega_a$, a time-dependent, dimensionless quantity that characterizes the apparent anisotropy, due to small temperature fluctuations in the primordial blackbody radiation:
\begin{equation}
\Omega_a \equiv \frac{\sigma^2 H^{-2}}{a^6 (t)} \quad \mbox{which scales as}
\quad \sim \frac{a^{2\epsilon}(t)}{a^6(t)};
\end{equation}
in this case, the patch size approaches zero faster than the horizon size.

A slow-twitch mechanism to explain the homogeneity, isotropy, and spatial flatness of the universe was proposed by~\citet{Khoury,Erickson}, in which the FLRW scale factor $a(\tau)$ decreases as a small power of the time parameter $\tau$, to model the slowing of the contraction of the universe:
\begin{equation}
   \lim_{\tau \to 0^-} a(\tau) \propto  (-\tau)^{1/\epsilon} \quad \mbox{where} \quad \epsilon \equiv \frac{3}{2} \bigl(1 + \frac{p}{\rho} \bigr) >> 3 \, ;
\end{equation}
in this equation, $\epsilon$ represents the equation of state, $p$ is the pressure, and $\rho$ is the energy density of the dominant stress-energy component.  

In most inflationary models the energy density $\rho$ is approximately constant, leading to
an exponential expansion of the scale factor
\begin{equation}
    a(t) \propto e^{\chi t} \quad \mbox{where} \quad \chi = \sqrt{\frac{8 \pi}{3}G\rho} \, , 
\end{equation}
with the state that drove the inflation involving a scalar inflation field (false vacuum)  in a local (but not global) minimum of its potential energy function. And in order to avoid the randomness of the bubble formation when the false vacuum decay would produce disastrously large inhomogeneities, the inflaton potential needs not have either a local minimum or a gentle plateau. 

In a bouncing or cyclic cosmology based on slow contraction, the evolution connects to the hot expanding universe through a classically-described smooth transition (the ‘bounce’) and the conversion of energy driving slow contraction into hot matter and radiation~\citep{Ijjas2014,Ijjas2018,Ijjas2019}. 
Numerical relativity simulations thus determined the number of bounces to achieve causality, more precisely, that the scale factor must grow by a factor of nearly $e^{\cal N} = e^{60}$ (or ${\cal N} = 60$ ``e-folds'') before the horizon size grows to equal the patch size in the expanding phase~\citep{Ijjas2019}.

The solution of the branch-cut-type cosmology corresponds to the reciprocal of a complex  multi-valued function, the natural complex logarithm function $\ln[\beta(t)]$, which
corresponds to a helix-like superposition of cut-planes, the Riemann sheets, with an upper edge cut in the $n$-th plane joined with a lower edge of cut in the ($n + 1$)-th plane. $\ln[\beta(t)]$ maps an infinite number of  Riemann sheets onto horizontal strips, which represent in the branch-cut cosmology the time evolution of the time-dependent horizon sizes. The patch sizes in turn maps progressively the various branches of the $\ln[\beta(t)]$ function which are {\it glued} along the copies of each upper-half plane with their copies on the corresponding lower-half planes. In the branch-cut cosmology, the cosmic singularity is replaced by a family of Riemann sheets in which the scale factor shrinks to a finite critical size, --- the range of $\ln^{-1}[\beta(t)]$, associated to the cuts in the branch cut, shaped by the $\beta(t)$ function ---, well above the Planck length.  In the contraction phase, 
as the patch size decreases with a linear dependence on $\ln[\beta(t)]$, light travels through geodesics on each Riemann sheet,  circumventing continuously the branch-cut,  and although the horizon size scale with $\ln^{\epsilon}[\beta(t)]$, the length of the path to be traveled by light compensates for the scaling difference between the patch and horizon sizes. 

In these circumstances, causality between the
horizon-size and the patch-size can be achieved through the accumulation of branches in the transition region between the current state of the universe, 
with the local time of travel of light around each Hubble horizon 
and past events associated with the global time of evolution of the universe, 13.8 billion years ago. 

To calculate the number of branches, we consider that the Riemann leaves correspond backwards to primordial stages of Hubble horizons, infinitesimally separated and that light travels through the extreme geodesics paths of a conical helix spiral. In this sense, we need to calculate the total value of the linear length corresponding to a conical spiral with several evolutionary stages, where the radius $r$ is a continuous, monotonic and decreasing function of the angle.

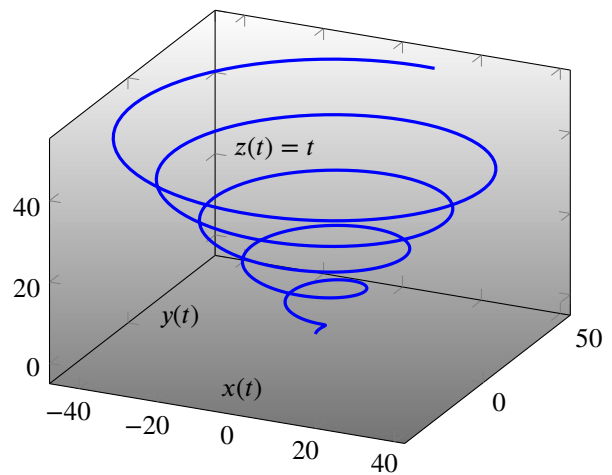
\begin{figure}
    \centering
    \begin{tikzpicture}
  \begin{axis}[
    axis background/.style =
      {shade, top color=white, bottom color=black!55}]
  \addplot3+ [
      domain  = 0:16*pi,
      samples = 400, samples y  = 0,
      mark    = none,
      very thick,
    ]
    ( {x*sin(0.2*pi*deg(x))},{x*cos(0.2*pi*deg(x)},{x});
\draw (0,-25) node[anchor=north east] {$x(t)$}
      (-27,0) node[anchor=east]{$y(t)$};        % want it near point (0, 2)
       \node[] at (-6,-10, 48) {$z(t) = t$};
  \end{axis}
\end{tikzpicture}
  \caption{PGF/TikZ vector graphics drawing of the conical spiral parameterized equation \ref{conical}. }
    \label{fig1}
\end{figure}

Assuming a conical spiral parameterized equation $\mathbf{r}(t)$, with parameter $t$, 
initial radius $r$ and slope $m =1$ (see Figure 1), 
\begin{equation}
\mathbf{r}(t)   =  \bigl(x(t) = r \cos(t), y(t) = r \sin(t), z(t) = t\bigr), \label{conical}
\end{equation}
with $n$ denoting the number of branches. 
Just as an illustration of the behavior of the equation \ref{conical}, the figure \ref{fig1} presents numerical values in relative scales and not absolute values.
%%%%%%%%%%%%%%%%%%%%%%%%%%%%%%%%%%%
%%%%%%%%%%%%%%%%%%%%%%%%%%%%%%%%%%%%%%%%%%%%%%%
The total linear length corresponds to 
\begin{eqnarray}
L   & = & \int_0^{2n\pi} \sqrt{(x^{\prime}(t))^2 + (y^{\prime}(t))^2 + (z^{\prime}(t))^2}dt 
 =  \sqrt{8} n\pi r \nonumber \\
& \to & 
n = \frac{L}{\sqrt{8} \pi r}. \label{n}
\end{eqnarray}
Now taking 
$L=$ age of the universe $\times$ light speed,  and as an extreme limit, for evaluation purposes, $r$ equal to the Planck scale, although it corresponds to the superposition region of classical and quantum effects, equation~(\ref{n}) gives 
\begin{equation} n = 1.055 \times 10^{61}.
\end{equation}
This is the number of `turns' or Riemann sheets or running-number of circulation sheets or even the number of branches in between the present state of the universe and the region where causality is achieved, more precisely, the cosmic microwave background radiation's "surface of last scatter". 
This result is comparable to 60 e-folds of contraction in the FLRW cosmic scale factor $a(t)$ to overcome causality achieved in the bouncing model~\citep{Ijjas2019}. 

\section{Summary and discussion}

The impossibility of packaging energy and entropy according to the Bekenstein criterion in a finite size makes the transition phase of the branch-cut cosmology very peculiar, imposing a topological leap between the two phases or a transition region similar to a wormhole, with space-time shaping itself topologically in the format of a helix-shape like around a branch-point~\citep{Zen2021b}, a topic that needs further investigation in the future. A quantum approach, more in line with the nature of this evolutionary domain~\citep{Hess2020}, is under investigation. 

Other aspects to be investigated refer to a more consistent evolution mapping of the cut-planes, the consequences of adopting a non-symmetric approach regarding the dimensionless thermodynamics connection $\epsilon(t)$, a different ordering of values of $\epsilon(t)$, and the role of dark matter in the evolution of the branch-cut universe. Cosmic Microwave Background (CMB) measurements by the Planck satellite~\citep{Aghanim2020} offer on the other hand an unprecedented opportunity to constrain and test the branch-cut model, particularly regarding the fluctuations associated to the energy density primordial spectrum, the seeds of all structures in the early universe. 

There are also questions regarding the multiverse content. For example, the conjuncture of primordial multi-universes immersed in a thermal bath, subjected to a contraction crunch, and about the thermal evolution consequences to our universe after decoupling. As well as ensuing questions, as the homogeneity of the primordial radiation and average density of matter as key elements for the formation of the structures observed today in the universe, among others. Evidently, most of these questions should not be limited to a classical description of the universe evolution, as their nature is intrinsically quantum. These topics are presently under investigation.

Throughout human history, in the writings of essayists, philosophers, or scientists, there is a recurring question about their ideas and conceptions about time, the flow of time and even, in a more philosophical sense, about eternity. In this particular, the word `mystery' as a qualifier of a difficulty inherent to the elucidation of this recurring theme is perpetuated over time. Branch-cut cosmology, in turn, lacks, despite some sketches made in the past, a formulation that addresses another mystery, a quantum mechanical approach required for physics at the Planck Scale, in which spacetime geometry is a quantum variable, through generalizations of usual quantum theory that incorporates spacetime alternatives, gauge degrees of freedom, and histories that move forward and backward in time~\citep{Hartle}. 

As a final message, we quote Jorge Luis Borges, renowned Argentine essayist, ``I do not pretend to know what time is (not even if it is a thing), but I guess that the course of time and time are a single mystery and not two." And he then continues, "Time is the substance of which I am made. Time is a river that snatches me away, but I am the river; it is a tiger that destroys me, but I am the tiger; it is a fire that consumes me, but I am the fire"~\citep{Borges} (translation of the authors).

\bibliography{Zen.bib}%
\end{document}